\documentclass[aps,prc,preprint,groupedaddress,preprintnumbers,
amsmath,amssymb]{revtex4}

\usepackage{graphicx}
\usepackage[figuresright]{rotating}

\begin{document}

\title{Electromagnetic reactions of few-body systems with the Lorentz 
       integral transform method}

\author{W. Leidemann}

\affiliation{Department of Physics, George Washington University,
        Washington DC 20053, USA \\and
         Istituto Nazionale di Fisica Nucleare, Gruppo Collegato di Trento, Italy} 
\thanks{On leave of absence from Dipartimento di Fisica, Universit\`a di Trento, 
        I-38050 Povo (Trento), Italy}
       


\begin{abstract}
Various  electromagnetic few-body break-up reactions into
the many-body continuum are calculated microscopically 
with the Lorentz integral transform
(LIT) method. 
For three- and four-body nuclei the nuclear Hamiltonian 
includes two- and three-nucleon
forces, while semirealistic interactions are used in case of six- and
seven-body systems. Comparisons 
with experimental data are discussed. In addition
various interesting aspects of the $^4$He photodisintegration are studied: 
investigation of a tetrahedrical symmetry of
$^4$He and a test of non-local nuclear force models via the induced two-body currents.
\end{abstract}

\maketitle

\section{The Lorentz integral transform method}

The LIT method was introduced about a decade ago \cite{ELO94}. It 
makes feasible calculations of observables, where many-body
continuum states are involved. In the LIT approach
explicit calculations of continuum state wave functions are avoided,
but nonetheless the continuum state interaction
is fully taken into account. This is achieved via the use of integral 
transforms with a formal reduction
of an A-body continuum state problem to an A-body 
bound-state-like problem \cite{Efros} much simpler to solve. 
The method 
can be applied to inclusive
and exclusive reactions; here we will only describe the inclusive case briefly. 

Inclusive observables are expressed in terms
of response functions 
\begin{equation} 
R(\omega) = \sum_f |\langle f| \hat O | 0 \rangle |^2 \delta(\omega-E_f+E_0) \,,
\end{equation}
where $|0/f\rangle$ and $E_{0/f}$ are wave functions and energies of ground
and final states, respectively, while the transition operator $\hat O$ 
defines the specific $R(\omega)$. 
The first step
in a LIT calculation consists in the solution of the equation  
\begin{equation}
(H -E_0 -\sigma_R +i\sigma_I) |\tilde\Psi \rangle = \hat O | 0 \rangle \,,
\end{equation}
where $H$ is the nuclear Hamiltonian and $\sigma_{R/I}$ denote free real
parameters. Due to the asymptotically vanishing ground state on the right-hand side
and the complex energy term on the left-hand side, $\tilde\Psi$ is localized
and thus eq.~(2) can be solved with bound-state methods. 
In a next step one determines
the norm  $\langle \tilde\Psi | \tilde\Psi\rangle$. Using closure 
one can show \cite{ELO94}
that it is related to an integral transform of $R(\omega)$ with a 
Lorentzian kernel, i.e. the LIT:
\begin{equation}
\langle \tilde\Psi | \tilde\Psi\rangle = 
\int d\omega {\frac {R(\omega)} {(\omega-\sigma_R)^2 + \sigma_I^2} } \,.
\end{equation}
Calculating the LIT for many values of $\sigma_R$ and
fixed $\sigma_I$ enables one to invert
the transform reliably in order to obtain  $R(\omega)$ (for 
inversion methods
see \cite{inv}).
For the solution of $\tilde\Psi$ we use expansions in hyperspherical
harmonics (HH). 
To improve the convergence of the expansion we use two different methods:
(i) CHH expansions with additional two-body correlation functions \cite{CHH}
and (ii) EIHH expansions with an HH effective interaction
\cite{EIHH}.

\section{RESULTS AND DISCUSSION}
In the following we discuss various inclusive reactions for nuclei with
up to 7 nucleons: $^3$H/$^3$He $(e,e')$ longitudinal form 
factors $R_L(\omega,q)$
(CHH calculations with realistic nuclear forces \cite{ELOT1,ELOT2}) 
and total photoabsorption
cross sections $\sigma_\gamma(\omega)$ of 4-, 6- and 7-body nuclei 
(EIHH calculations
with realistic nuclear force for A=4 \cite{DBBLO} and semirealistic 
NN potentials for A=6, 7 \cite{6body,7body}).
In case of $R_L$ we take 
the non-relativistic (n.r.) nuclear charge operator as transition operator $\hat O$,
but also include relativistic corrections (Darwin-Foldy and spin-orbit terms).
For the calculation of $\sigma_\gamma$ we use the unretarded dipole 
approximation, 
which is excellent 
for $\omega < 50$ MeV as was shown in two- and three-nucleon studies. 

\begin{figure}[htb]
\begin{minipage}[t]{80mm}
\resizebox*{7cm}{10cm}{\includegraphics*[angle=0]{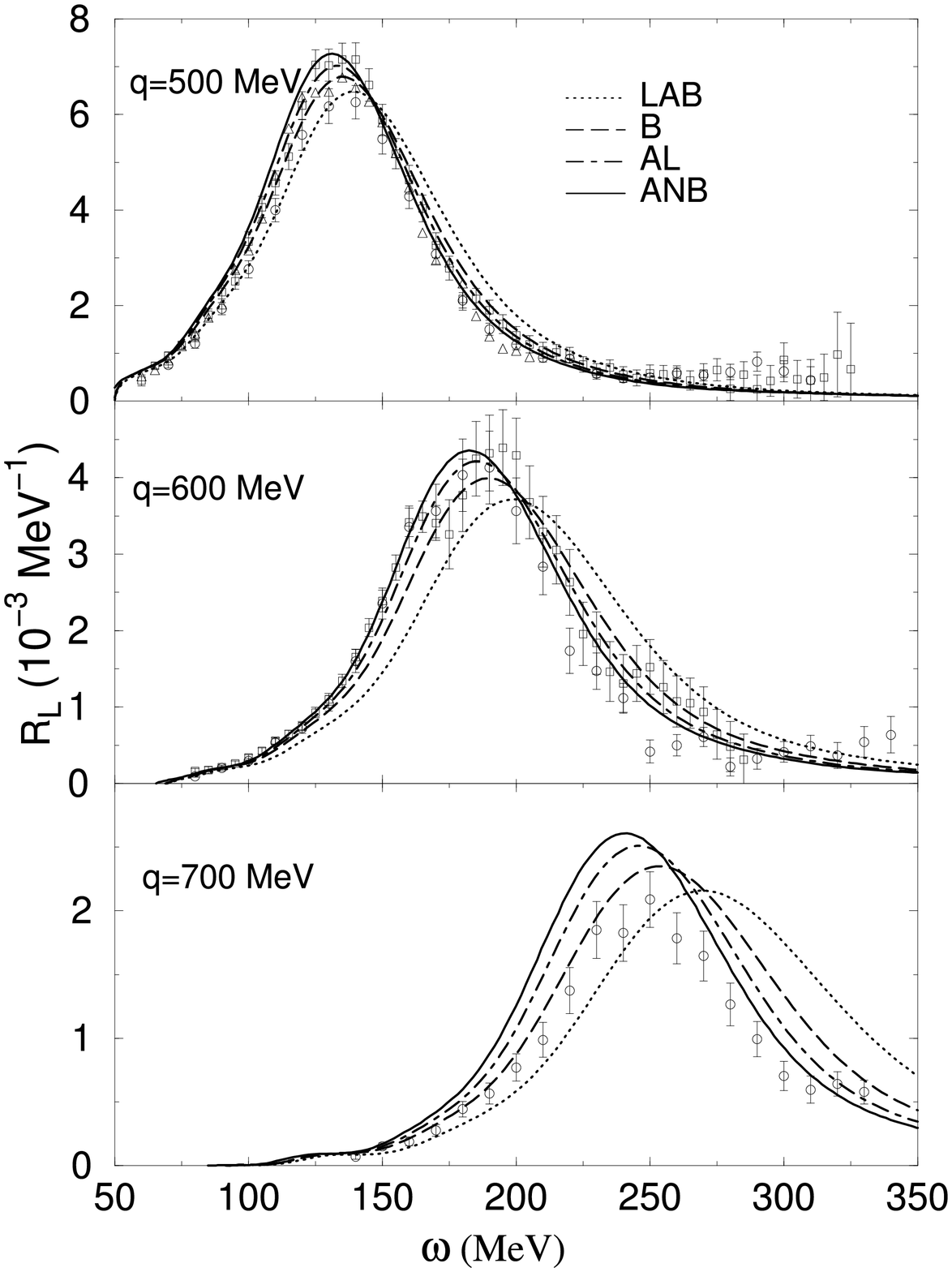}}
\caption{$R_L(\omega,q)$ of $^3$He at various $q$ calculated
in different frames (see text); experimental data from \cite{Saclay} (triangles)
,
\cite{Bates} (squares), \cite{Carlson} (circles).}
\end{minipage}
\label{figure1}
\hspace{\fill}
\begin{minipage}[t]{75mm}
\resizebox*{7cm}{10cm}{\includegraphics[angle=0]{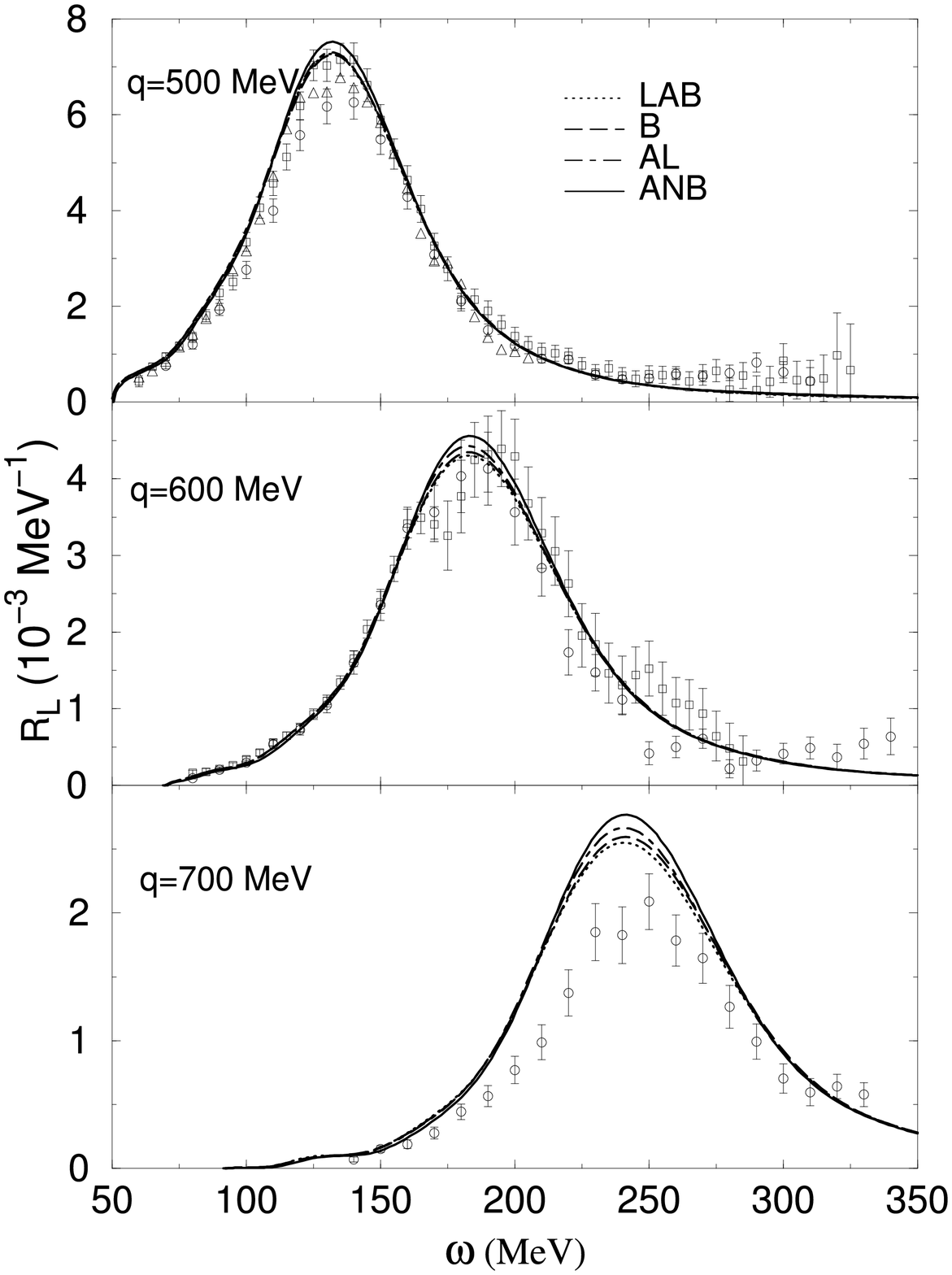}}
\caption{As Fig.~1 but using two-body relativistic kinematics for the final
state energy as discussed in the text.
}
\end{minipage}
\label{figure2}
\end{figure}
In \cite{ELOT1} we calculated $R_L(\omega,q)$ of $^3$H/$^3$He
at 250 MeV/c $\le q \le 500$ MeV/c with  
the AV18 NN and the UIX 3N potentials. We found that
the 3N-force
reduces the quasielastic (q.e.) peak height, which improves the 
agreement with experimental data in case of $^3$He, but worsens
the agreement in the triton case. On the other hand it should be
noted that the experimental situation
is more settled for $^3$He, since there one has two independent  
data sets and not only one as in the $^3$H case. 
It is interesting to study the validity of a n.r. $R_L$ calculation 
for increasing
$q$. A kind of minimal check is made by calculating the
$R_L^{\rm fr}(\omega_{\rm fr},q_{\rm fr})$ of different reference frames and by 
transforming them relativistically correctly to the laboratory
system (note our definition $R_L(\omega,q) \equiv 
R_L^{\rm lab}(\omega_{\rm lab},q_{\rm lab})$) according to 
\begin{equation}
R_L(\omega,q) = {\frac {q^2} {q_{\rm fr}^2} } {\frac {E_T^{\rm fr}} {M_T} }
R_L^{\rm fr}(\omega_{\rm fr},q_{\rm fr}) \,,
\end{equation}
where $E_T^{\rm fr}$ and $M_T$ are energy and mass of the 
target nucleus. 
In Fig.~1 we show results for four different frames
\cite{ELOT2}, which are defined via the target nucleus 
initial momentum
${\bf P}_T^{\rm fr}$: ${\bf P}_T^{\rm lab}$=0,
${\bf P}_T^{\rm B}=-{\bf q}/2$, ${\bf P}_T^{\rm AL}=-{\bf q}$,
${\bf P}_T^{\rm ANB}=-3{\bf q}/2$. It is readily seen 
that one obtains more and more frame
dependent results with growing $q$.
As shown in \cite{ELOT2} the frame dependence can be drastically 
reduced if one imposes
q.e. kinematics 
and takes the relativistic relative
momentum between knocked-out nucleon and residual two-body system
as input in the calculation. As seen in Fig.~2 one
finds almost frame independent results, which show a very good agreement with
experimental data at $q$ = 500 and 600 MeV/c, while at 700 MeV/c the experimental 
peak height
is considerably lower than the theoretical one. We would like to point out
that the ANB result of Fig.~1 lies within 
the band of almost frame independent
results of Fig.~2. Thus we consider the ANB frame as the optimal frame for
a n.r. calculation. It leads to a proper description of
the q.e. peak region, however, with no need to assume q.e. kinematics
and thus its validity is not restricted to the peak region
(further explanations why the ANB system is preferable 
are given in \cite{ELOT2}). 

\begin{figure}[htb]
\begin{minipage}[t]{80mm}
\resizebox*{7cm}{8cm}{\includegraphics*[angle=0]{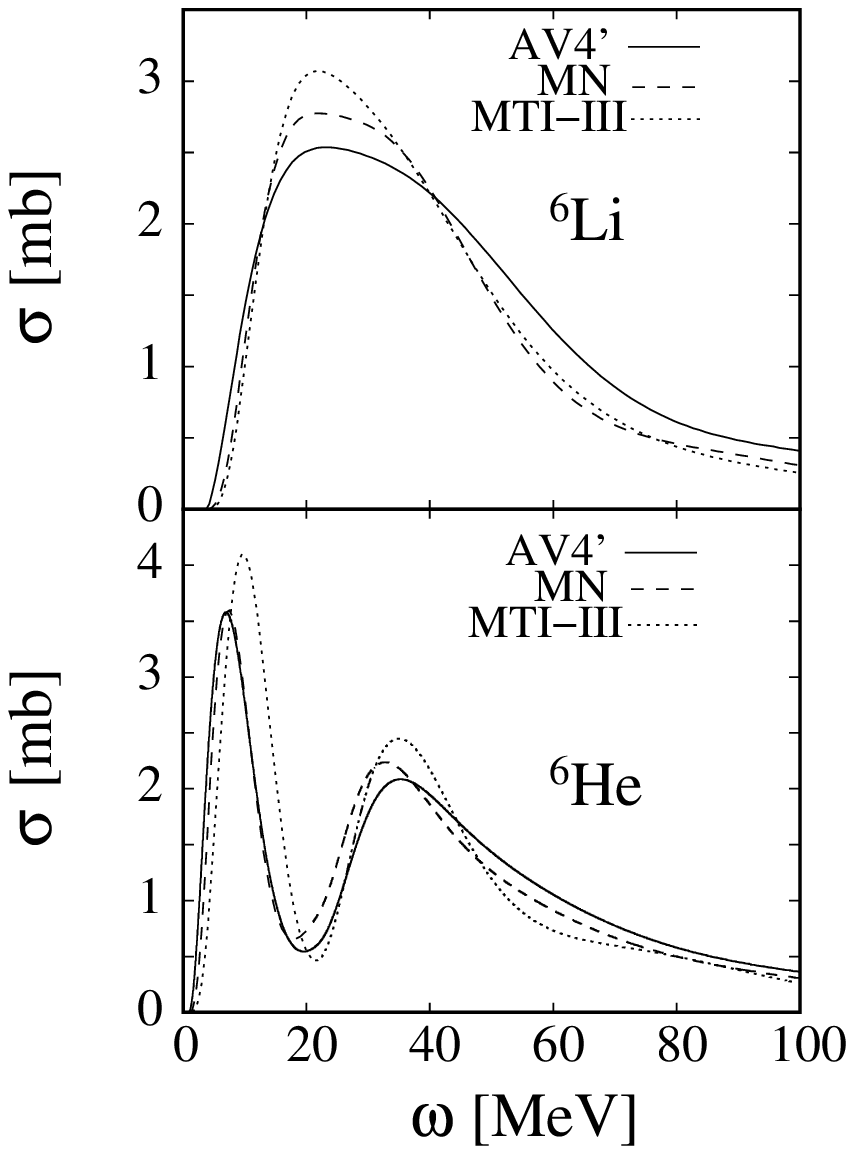}}
\caption{Total photoabsorption cross sections of $^6$Li and $^6$He calculated
with various semirealistic potential models.}
\end{minipage}
\label{figure3}
\hspace{\fill}
\begin{minipage}[t]{75mm}
\resizebox*{8cm}{8cm}{\includegraphics[angle=0]{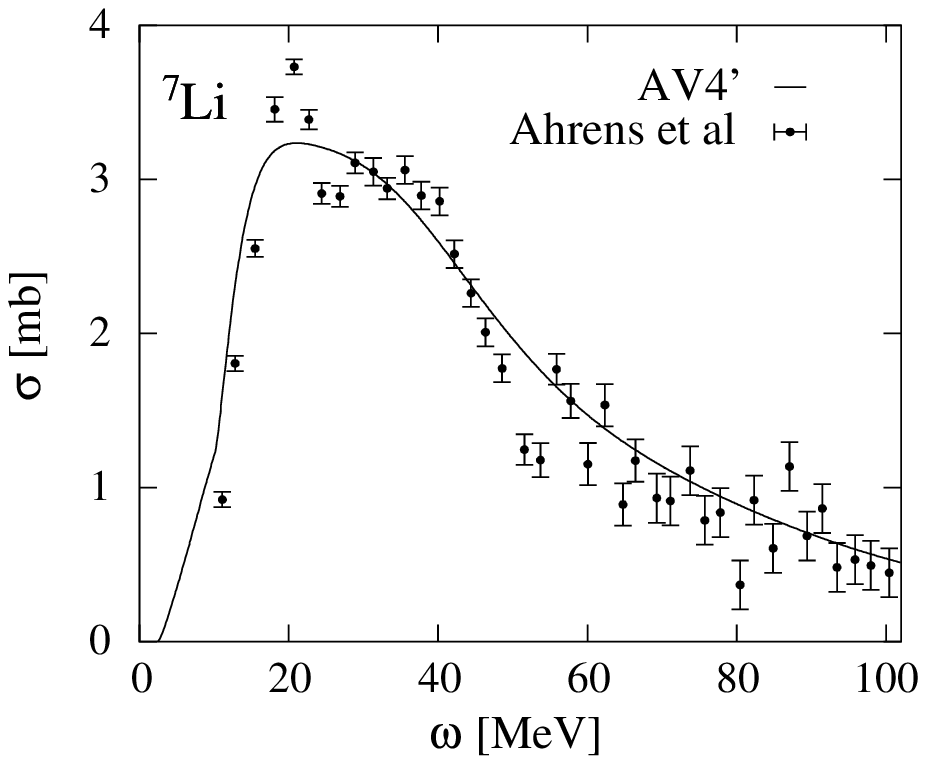}}
\caption{$^7$Li total photoabsorption cross section calculated with the
AV4' potential; experimental data from \cite{Ahrens}.
}
\end{minipage}
\label{figure4}
\end{figure}
Now we turn to nuclear photodisintegration and first discuss A$\ge 6$ nuclei. 
In Fig.~3  we show $\sigma_\gamma(\omega)$
of $^6$He and $^6$Li calculated 
with various semirealistic 
NN potential models: Malfliet-Tjon
(MT), Minnesota (MN), and Argonne V4' (AV4') \cite{6body}.
Particularly interesting are the $^6$He results, they show two separate
peaks. In a many-body picture they correspond to a soft mode at low energies, 
where
the surface neutrons oscillate against the $\alpha$-core, and a Gamov-Teller
mode at higher energies, where the neutrons oscillate against the protons.
It is worthwile to point out that such a double peak structure comes out
also in our microscopic
few-body calculation.
For a comparison with experiment we refer to \cite{6body}, 
here
we show a comparison with data for  the $^7$Li case \cite{7body} 
(see Fig.~4). It is evident that there is a rather good 
agreement between theory and experiment,
which also shows that with the LIT calculation
one has a reliable control of the seven-body continuum.

The experimental study of the $^4$He photodisintegration has a rather
long history. First experiments have been carried out
about 50 years ago. Much work was devoted to the two low-energy
two-body break-up channels of $^3$H-$p$ and $^3$He-$n$. The results
were rather controversial showing either a very pronouned giant dipole
peak or a rather flat structure. In the '80s and '90s 
data seemed to converge to a less pronounced peak 
\cite{Berman,Feldman}, which, however, is at variance with a following
determination via Compton scattering \cite{Wells}. In one of our first
LIT applications we calculated $\sigma_\gamma(\omega)$  of $^4$He 
using semirealistic potential 
models \cite{ELO97}. In contrast
to the experiments of \cite{Berman,Feldman} we found
a very pronounced giant dipole peak, which also revived experimental
activities (see \cite{Nilsson,Shima}). A very important further theoretical clarification
comes from our recent calculation
of $\sigma_\gamma(^4$He) with a realistic nuclear force (AV18+UIX)
\cite{DBBLO}. In Fig.~5 one sees that the 3N-force leads to
a reduction of the peak height, but that the full AV18+UIX result shows 
a pronounced peak and thus confirms the findings of \cite{ELO97}.  
In comparison to experiment one observes a rather good agreement
with the data of \cite{Nilsson} (note: measured 
$(\gamma,n)$ cross section is doubled assuming that it is about equal to the
$(\gamma,p)$ cross section; only statistical errors are shown), while the data
of \cite{Shima} are much lower and do not exhibit any low-energy peak.
\begin{figure}[htb]
\begin{minipage}[t]{87mm}
\resizebox*{8cm}{8cm}{\includegraphics*[angle=0]{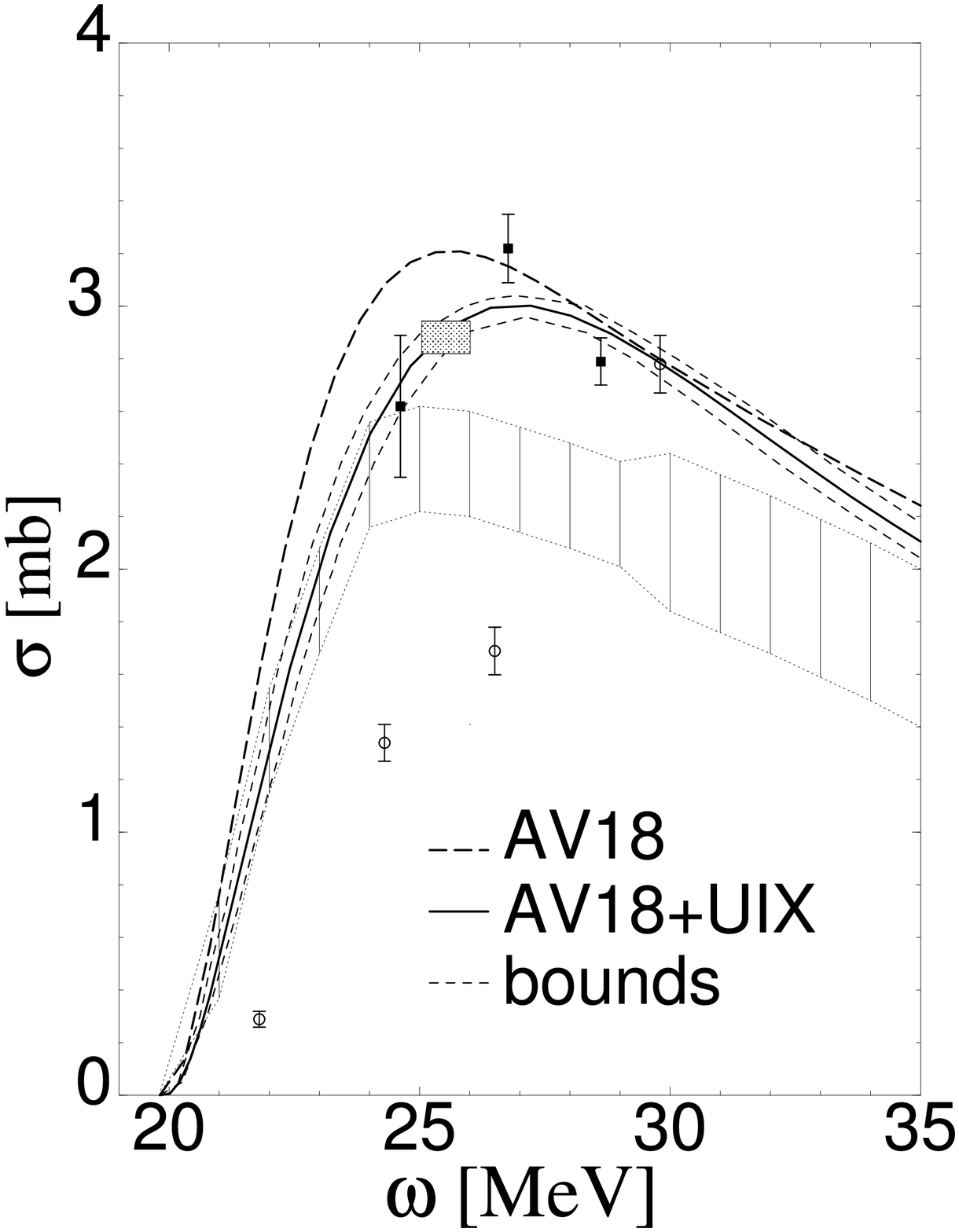}}
\caption{$^4$He total photoabsorption cross section with AV18+UIX 
(with upper and lower bounds due to uncertainties 
of HH expansion convergence) and AV18 potentials; experimental data
(see also text): area between
dotted lines \cite{Berman,Feldman}, dotted box \cite{Wells}, squares
\cite{Nilsson}, and circles \cite{Shima}.}
\end{minipage}
\label{figure5}
\hspace{\fill}
\begin{minipage}[t]{65mm}
\resizebox*{7cm}{7cm}{\includegraphics*[angle=0]{wl_fig6.eps}}
\caption{As Fig.~5 but calculated with AV18+UIX,
JISP and UCOM potentials.}
\end{minipage}
\label{figure6}
\end{figure}

A calculation of $\sigma_\gamma$ in unretarded dipole approximation has 
the great advantage
that also the dominant part of the meson exchange current (MEC) contribution is 
considered (Siegert theorem). Thus $\sigma_\gamma$
is an ideal testground for nuclear potential models:
one works with a very simple transition operator $\hat O$, but
nonetheless includes an important MEC contribution. Compared
to purely hadronic reactions one obtains additional valuable information about
a given nuclear force model. We take advantage of this fact and test
potentials \cite{test,Sonia} designed for the use in more complex nuclei (JISP models
\cite{JISP}, UCOM \cite{UCOM}), which on the one hand describe NN scattering 
data, but on the other hand also
binding energies of nuclei with $A>2$ (see also contribution of G. Orlandini
et al.). In Fig.~6 we show results for the $\sigma_\gamma$ of $^4$He 
for these potentials in comparison to the AV18+UIX case. One sees
that the peaks are a bit lower and also
shifted somewhat towards higher energy for JISP and UCOM interactions. 
Beyond the peak there is a rather
good agreement between UCOM and AV18+UIX results, while the JISP6 potential
leads to a considerably higher cross section. Due to the lack of precise
experimental data one cannot rule out any of these potential models.

Additional information on the electromagnetic
structure of nuclei can be obtained from photonuclear sum rules.  
In \cite{sumrules} we have considered various
$^4$He sum rules with the AV18+UIX potential, but here we only discuss
the $^4$He bremsstrahlungs sum rule
\begin{equation}
\Sigma_{\rm BSR} = \int_{\omega_{th}}^\infty d\omega \, {\frac {\sigma_\gamma^{E1UR}}
{\omega} } \,,
\end{equation}
where $\sigma^{E1UR}$ is $\sigma_\gamma$ in unretarded
dipole approximation. One also has \cite{sumrules}
\begin{equation}
{\frac {\Sigma_{\rm BSR}}{g}} = Z^2 \langle r_p^2 \rangle  - {\frac {Z(Z-1)}{2}}
\langle r_{pp}^2 ) 
 = N^2 \langle r_n^2 \rangle  - {\frac {N(N-1)}{2}}
\langle r_{nn}^2 ) 
 =  {\frac {NZ}{2}} \Bigl( \langle r_{np}^2 \rangle  - \langle r_n^2 \rangle
- \langle r_p^2 \rangle \Bigr) 
\end{equation}
with $g=4\pi^2\alpha/3$ and 
where $\langle r_{p/n}^2 \rangle$ is the mean square proton/neutron radius
and $\langle r_{pp/nn/np}^2 \rangle$ is the mean square distance between
nucleons in a 
pp/nn/np-pair. Neglecting the tiny isospin breaking part of the NN interaction,
for $^4$He one has $\langle r_p^2 \rangle = \langle r_n^2 \rangle$. In case of the 
AV18+UIX potential
one finds $\langle r_{n/p}^2 \rangle$=2.04 fm$^2$, which is in agreement with the
$^4$He experimental charge radius. Thus an additional 
determination of $\Sigma_{\rm BSR}$
leads to the various NN mean square distances. Our evaluation
for AV18+UIX yields a $\Sigma_{\rm BSR}$ value of 2.410 mb leading to
5.67 fm$^2$ ($ \langle r_{pp/nn}^2 \rangle$) and 5.34 fm$^2$ 
($ \langle r_{np}^2 \rangle$).
The difference between the two values is caused by the isospin dependence of the nuclear force
($pp/nn$-pairs: isospin T=1, $np$-pairs: T=0,1).
It is interesting to note that the ratios $ \langle r_{pp}^2 \rangle
/ \langle r_{p}^2 \rangle$= 2.78 and $ \langle r_{np}^2 \rangle /
\langle r_{p}^2 \rangle$= 2.62 are very close to the value for a tetrahedrical
symmetry of 2.67 \cite{sumrules}. Thus one may conclude that the $^4$He
two-body density reflects to a large extent such a symmetry.

It follows a short summary of this presentation.
It has been shown that the LIT method is a very powerful tool: 
a continuum state problem is reduced to a bound-state-like problem. It enables one
to make ab inito calculations of reactions with light
nuclei into the many-body continuum. Among the discussed applications
is the first cross section calculation of a four-body reaction into the
four-body continuum ($^4$He total photoabsorption cross section)
evaluated with a realistic nuclear force (AV18+UIX).
The results  
show a pronounced giant resonance peak, while the
experimental data do not lead to a unique picture and hence further 
experimental investigations
are certainly necessary.

\end{document}